\documentstyle[12pt,twoside]{article}
\newcommand{\hst}{\hspace{-0.3cm}}
\newcommand{\hsp}{\hspace{0.2cm}}

\newcommand{\beqn}{\begin{equation}}
\newcommand{\eeqn}{\end{equation}}
\newcommand{\barr}[1]{\begin{array}{#1}}
\newcommand{\earr}{\end{array}}
\newcommand{\beqna}{\begin{eqnarray}}
\newcommand{\eeqna}{\end{eqnarray}}
\newcommand{\btablec}{\begin{table} \begin{center}}
\newcommand{\etablec}{\end{center} \end{table}}

\begin{document}
\title{
\small \rm \begin{flushright} \small{hep-ph/9412301}\\
\small{RAL-94-122}\\
\small{OUTP-94-35P} \end{flushright}
\vspace{2cm}
\LARGE \bf The photoproduction of hybrid mesons from CEBAF to HERA
\vspace{0.8cm} }
\author{Frank E. Close\thanks{E-mail : fec@v2.rl.ac.uk} \\
{\small \em Particle Theory, Rutherford-Appleton Laboratory, Chilton,
Didcot OX11 0QX, UK} \\ \\
% \thanks{} directly next to footnote point creates a footnote
Philip R. Page\thanks{E-mail : p.page@physics.oxford.ac.uk} \\
{\small \em Theoretical Physics, University of Oxford, 1 Keble Road, Oxford OX1
3NP, UK}  \\  \\}
% no date \date{}
\date{December 1994 \vspace{1.5cm}}

\begin{center}
\maketitle

\begin{abstract}

Decay widths, branching ratios and production dynamics of some
recently discovered $J^{PC} = 1^{-+}, 0^{-+}$,$1^{--}$ and $2^{-+}$
mesons are found
to be in remarkable agreement with the predicted properties of
hybrid mesons. We propose tests for this new dynamics,
emphasise the critical role of $\pi b_1$ or $\pi h_1$ decay channels
in discriminating hybrids from conventional states, and suggest that
photoproduction may offer special opportunities for isolation and
confirmation of hybrids.

\end{abstract}
\end{center}

%%%%%%%%%%%%%%%%%%%%%%%%%%%%%%%%%%%%%%%%%%%%%%%%%%%%%%%%%%%%%%%%%%%%%%%
\newpage

There is a substantial area in the Standard Model of particles and interactions
where we remain
fundamentally ignorant: while the gluon degrees of freedom expressed
in the Lagrangian of Quantum Chromodynamics
have been established beyond doubt in high momentum data,
their dynamics in the strongly interacting limit epitomised by hadron
spectroscopy are quite obscure.
It is possible that this is about to change \cite{amsler94}.

 For the first time there is a  candidate scalar glueball
 \cite{anisov,kirk,bugg} whose mass
$1.5 \sim 1.6$ GeV is consistent with the prediction of $1550 \pm 50$ MeV from
lattice studies of QCD \cite{lattice}. Furthermore, its production and decay
properties in a range of experiments
fit more naturally with those of a gluonic excitation than
of a conventional quark - antiquark ($q\bar{q}$)
 nonet state.
If these are the first hints of gluonic excitation at a mass scale
anticipated by theory, then this raises the possibility that in the
1.5 - 2 GeV mass range there is a rich spectroscopy of
``hybrid" states where the gluonic fields or ``flux--tubes"
are excited in the presence
of coloured quark sources\cite{barnes83,sharpe83,paton85}.
In this letter we present evidence supporting this notion and
propose ways of testing the hypothesis in forthcoming experiments.

It is well known that hybrid mesons can have quantum numbers for spin
parity and charge conjugation ($J^{PC}$) in combinations such as $0^{+-},
1^{-+}, 2^{+-}$ etc. which are unavailable to conventional mesons and as
such provide a potentially sharp signature for hybrids. A central theme
in this letter will be to note that even when hybrid and conventional mesons
have the {\bf same} $J^{PC}$ quantum numbers, they may still be distinguished.
The essential reason is that although
superficially identical in their overall quantum numbers, the two states have
different internal structures which give rise to
characteristic selection rules\cite{paton85,cp94}. Decays into $\pi b_1$ or
$\pi h_1$ are pivotal here: we illustrate this
for the particular case of
$1^{--}$ states\cite{cp94} such as may be produced in $e^+e^-$ annihilation or
diffractive photoproduction.

\vskip 0.1in

\noindent Specifically:

(i) If $q\bar{q}$ in either hybrid or conventional mesons
are in a net spin singlet configuration then the dynamics of the flux-tube
forbids decay into final states consisting only of spin singlet mesons.

For $J^{PC}=1^{--}$ states this selection rule
distinguishes between conventional vector mesons
which are $^3S_1$ or $^3D_1$ states and hybrid vector mesons where the
$q\bar{q}$ are coupled to a spin singlet. This implies that in the decays of
hybrid $\rho$, the channel
$\pi h_1$ is forbidden whereas $\pi a_1$ is allowed and
that $\pi b_1$ is analogously suppressed for hybrid $\omega$ decays; this
is quite opposite to the case of $^3L_1$ conventional mesons where the
$\pi a_1$ channel is relatively suppressed and $\pi h_1$ or $\pi b_1$
are allowed\cite{busetto,kokoski87}. The extensive analysis of data
in ref.\cite{don2} revealed the clear presence of $\rho(1460)$\cite{pdg94}
 with a strong $\pi a_1$ mode but no sign of $\pi h_1$,
in accord with the hybrid situation. Furthermore, ref.\cite{don2}
finds evidence for $\omega (1440)$ with no visible
decays into $\pi b_1$ which again is in significant contrast to the
expectations
for conventional $q\bar{q}$ $(^3S_1$ or $^3D_1$) initial states and in accord
with the hybrid configuration.

(ii) The dynamics of the excited flux-tube in the {\bf hybrid} state
suppresses the decay to mesons where the $q\bar{q}$ are $^3S_1$ or $^1S_0$
$``L=0"$ states. The preferred decay channels are to ($L=0$) + ($L=1$)
pairs\cite{paton85,kokoski85}. Thus
in the decays of hybrid $\rho \rightarrow 4\pi$ the $\pi a_1$ content is
predicted to be dominant and the $\rho \rho$ to be absent. The
analysis of ref.\cite{don2} finds such a pattern for $\rho(1460)$.

(iii) The selection rule forbidding ($L=0$) + ($L=0$) final states
no longer operates if the internal structure or size of the two ($L=0$) states
differ\cite{paton85,pene}.
 Thus, for example, decays to $\pi + \rho$, $\pi + \omega$ or $K + K^*$
may be significant in some cases\cite{cp94}. Though still suppressed
relative to the dominant pathway, (ii) above, this too
is the case for $\rho(1460)$.
It is also interesting to note that for a hybrid
$\omega (1440)$, the preferred $(L=0) + (L=1)$ decay paths are predicted to
be kinematically suppressed leaving the $\pi \rho$ and possibly $\eta \omega$
 as dominant decays\cite{cp94}.

(iv) In the case of {\it production}, where
an exchanged $\pi, \rho$ or $\omega$ is involved, it is possible that the
strength could be significant at least to the extent that the
exchanged off mass-shell state may have different structure to the incident
on-shell beam particle\cite{thankdon}.
 This may be particularly pronounced in the case
of {\bf photoproduction} where couplings to $\rho \omega$ or $\rho \pi$,
 which were suppressed in ref.\cite{cp94},
could be considerable when the $\rho$ is effectively replaced
by a photon and the $\omega$ or $\pi$ is exchanged. As we shall
discuss later, this may explain the production of the candidate exotic
$J^{PC}=1^{-+}$ (ref.\cite{lee94}) and a variety of anomalous
signals in photoproduction.

According to the flux-tube model, the $J^{PC}$ quantum numbers of the
 lowest lying hybrid mesons may be divided into two classes :

(a) $0^{-+},1^{+-},1^{--},1^{++},2^{-+}$
(deemed ``conventional" in that they can also
be shared by standard $q\bar{q}$ states), and

(b) $0^{+-},1^{-+},2^{+-}$ (deemed ``exotic").

In the hybrid $1^{--}$ and $1^{++}$ the  $q\bar{q}$ are coupled to a spin
singlet; all other $J^{PC}$ are spin triplets.

If the $\rho(1460)$ has signposted the existence of the
vector hybrid nonet, then we
need to establish which of the other seven multiplets should also be visible.
States whose couplings are predicted to
be strong, with highly visible decay channels and moderate widths relative
to the $\rho$ candidate, {\it must} be seen if hybrids are to be established.
Conversely, channels where
no signals are seen should be those whose signals are predicted in refs.
\cite{paton85,cp94} to be weak. We shall make the case
that these criteria do appear to be realised in the data.

\vskip 0.2in

\noindent {\underline{Exotic $J^{PC}$}}

\vskip 0.1in

The most obvious signature for a hybrid meson is the appearance
of a flavoured state with an exotic combination for $J^{PC}$.
It was noted in ref.\cite{kokoski85} and confirmed in ref.\cite{cp94}
that the $0^{+-}$ width is predicted to be over 1500 MeV thereby rendering
the state effectively invisible.
The $2^{+-}$ is also predicted to be very broad and hard to see
if its mass is $\geq$ 1.9 GeV.
The best opportunity for isolating exotic hybrids appears
to be in the $1^{-+}$ wave where, for the I=1 state with mass around 2 GeV,
partial widths are typically

\beqn
\label{bnlwidth}
 \pi b_1 : \pi f_1 : \pi \rho \;
= \; 170 \; MeV : 60 \; MeV : 10 \; MeV
\eeqn
The narrow $f_1(1285)$ provides a useful tag for the
$1^{-+} \rightarrow \pi f_1$ and ref.\cite{lee94} has recently reported a
signal
in $\pi^- p \rightarrow (\pi f_1) p$ at around 2.0 GeV
that appears to have a resonant phase though they admit that
more data are required for a firm conclusion.

The partial widths for $\pi b_1$ and $\pi f_1$ in eqn.(1)
are similar to those predicted
in ref. \cite{kokoski85} but we note
also the possibility that the $\pi \rho$ channel is
not negligible relative to the signal channel
$\pi f_1$ \cite{cp94}. This may be important in view of the
puzzle, commented upon in ref. \cite{lee94}, that
significant $\pi + \rho$ coupling may be present in
the production mechanism. In view of our analysis in ref.\cite{cp94}
and property
(iv) above, it is clear that the latter coupling may be
significant in $\pi p \rightarrow 1^{-+} p$.
We shall comment on the possible photoproduction of this state later.

\vskip 0.2in
\noindent {\underline{Conventional $J^{PC} $}}
\nopagebreak
\vskip 0.1in

Previous discussions of the dynamics of hybrids in flux-tube models have
been limited to exotic $J^{PC}$\cite{kokoski85}; the
first calculation of the widths and branching ratios of hybrid
mesons with conventional quantum numbers is in ref.\cite{cp94}. This predicts
that for hybrids at $\sim 2$ GeV made from $u,d$
flavoured quarks the $1^{+-},1^{++}$
states are over 500 MeV wide in both I=0,1 states; by contrast the
$0^{-+},2^{-+}$ and the
$1^{--}$ are predicted to be potentially accessible.
Hence we predict that the most visible $J^{PC}$ states are those that
coincide with the lowest lying hybrids of the bag model, namely
$1^{--}; {(0,1,2)}^{-+}$ \cite{barnes83,sharpe83}.
Moreover, the $q\bar{q}$ spin content is identical in the two models.
It is therefore interesting that each of these  $J^{PC}$ combinations
shows rather clear
signals with features characteristic of hybrid dynamics and which do not
fit naturally into a tidy $q\bar{q}$ conventional classification.

\vskip 0.1in
We have already mentioned the $1^{--}$ which motivated this analysis: a
more detailed discussion may be found in ref.\cite{cp94}.

\vskip 0.1in

Turning to the $0^{-+}$ wave, we note that the VES Collaboration
at Protvino sees an enigmatic
and clear $0^{-+}$ signal in diffractive production with 37 GeV
incident pions on beryllium \cite{ves}. They study the channels
$\pi^- N \rightarrow \pi^- \pi^+ \pi^- N; \; \pi^- K^+ K^- N$
and see a resonant signal $M \approx 1790$ MeV, $\Gamma \approx 200$ MeV
in the $(L=0)$ + $(L=1)$ $\bar{q}q$
 channels $\pi^- + f_0; \; K^- + K^*_0, \; K {( K \pi )}_S $ with no
corresponding strong signal in the kinematically allowed $L=0$ two body
channels $\pi + \rho; \; K + K^*$:

\beqn
\frac{0^{-+} \rightarrow \pi^- \rho^0}{0^{-+} \rightarrow\pi^- f_0(1300)}
\;  < \; 0.3 \; (95 \% \: C.L.)
\eeqn

\beqn
\frac{0^{-+} \rightarrow K^- K^*}{0^{-+} \rightarrow (K^-K^+ \pi)_S} \;
 < \; 0.1 \; (95 \% \: C.L.)
\eeqn
The width and large couplings to kaons are both surprising if this
were the second radial excitation of the pion (the first radial
excitation is seen as a broad enhancement in accord with expectations).
Furthermore, the preference for decay into $(L=0) + (L=1)$
mesons at the expense of $L=0$ pairs is in
accord with expectations for hybrids.

The resonance also appears to couple strongly to
the enigmatic $f_0(980)$:

\beqn
\frac{0^{-+} \rightarrow \pi^- f_0(980)}{0^{-+} \rightarrow \pi^- f_0(1300)}
\;  = \; 0.9 \pm 0.1
\eeqn
As noted in ref. \cite{ves} this is an unexpectedly high value since the
$f_0(980)$ has a small width and strong coupling to strangeness while the
$f_0(1300)$ is a broad object coupled mainly to non-strange quarks.
However, this may be natural for a hybrid at this mass for the following
reason.
The $KK^*_0$ channel would be dominant
(table 3 of ref.\cite{cp94}) but for the fact that it
 is kinematically suppressed; however,
the $\sim 300$ MeV width of the $K^*_0(1430)$ will feed
the $(KK\pi)_S$ (as observed \cite{ves}) and help to feed the channel
$\pi f_0(980)$ through the strong affinity of $K\bar{K} \rightarrow f_0(980)$.
Thus the overall expectations for hybrid $0^{-+}$ are in line with
the data of ref.\cite{ves}. Important tests are now that there should be
a measureable decay to the $\pi \rho$ channel with only a small
$\pi f_2$ or $KK^*$ branching ratio.
%\vskip 0.2in
%{\underline{$2^{-+} $}}

\vskip 0.1in

This leaves us with the $2^{-+}$.

\vskip 0.1in

There are clear signals of unexplained activity in the
$2^{-+}$ wave in several experiments for which a hybrid interpretation
may offer advantages.

Historically the ACCMOR Collaboration
\cite{accmor} noted a $2^{-+}$ structure around
1.8 GeV, coupled to $\pi f_2$, and too near to the  $ \pi_2(1670)$
for these to be the $1^1D_2$ and  $2^1D_2$ (radial excitation) of conventional
quarkonium. Chanowitz and Sharpe suggested \cite{sharpe83} that the
$1.8$ GeV state might set the mass scale for hybrid excitations.
This structure is tantalisingly
similar to sightings of a possible $2^{-+}$ (or even $1^{-+}$, see
below) at $1.77$ GeV, width 100-200 MeV in photoproduction via
$\pi$ exchange \cite{condo} and coupled to $\pi \rho$ and $\pi f_2$.
An earlier low energy photoproduction experiment \cite{eisenberg}
also shows a clear structure at 1.7 - 1.9 GeV though its quantum numbers are
not
identified; we
note that hybrid $2^{-+}, 1^{-+}$ are both favourably photoproduced
according to the table 7,8 in ref.\cite{cp94} {\it via}
$\pi$ exchange or $\omega$
exchange.
There are also some indications of a doubling of states in the $I=0$
 $\eta \pi^0 \pi^0$
channel where the Crystal Barrel at LEAR\cite{cooper} finds
both $\eta_2(1650)$ (which is probably the partner of $\pi_2(1670)$)
and also a candidate $\eta_2(1850)$ decaying into $f_2 \eta$ (unlikely to be
$s\bar{s}$).

If these various experiments are heralding activity in the $2^{-+}$
wave, a search for the $\pi b_1$
decay channel \cite{cp94} becomes pivotal. This follows once again from
the selection rule forbidding the decay of a spin singlet meson into
pairs of spin singlets: this prevents the decay of
$^1D_2 (\pi_2) \rightarrow b_1 \pi$ whereas this channel is allowed and
potentially significant for $\pi_2^{hybrid}$ \cite{cp94}.

In this context, the results of ref.\cite{atkinson} are interesting.
They studied
$\gamma p \rightarrow (b_1 \pi) p$ at 25-50 GeV incident energy with the
specific intention of seeking hybrids.
At that time the only calculations of hybrid branching ratios in the literature
were for the {\bf exotic} $J^{PC}$ states
and ref.\cite{atkinson} assumed them to be
a guide to the non-exotic cases also. Our calculations \cite{cp94}
show that this assumption is not generally true. In particular, as noted above,
 the selection rule forbids the
$b_1 \pi$ decay modes for the case of hybrid
$\omega$ and conventional $^1D_2$ ($\pi_2$) quarkonium while allowing
it for conventional ($^3S_1$ or $^3D_1$) $\omega$ or hybrid $\pi_2$
respectively. As the $b_1 \pi$ is the trigger channel in ref.\cite{atkinson}
several of the conclusions of that experiment merit re-examination.
In particular, if one insists on diffractive
production then any signals are {\bf not} hybrid $1^{--}$; conversely, if one
allows also $\omega$ exchange (which is not negligible
at 25-50 GeV), then either (or both!) hybrid $2^{-+}, 1^{-+}$ production
can feed the $b_1 \pi$ signal but conventional $^1D_2$ quarkonium is forbidden
in this mode. $\omega$ exchange also can feed $1^{++}$: a $2^3P_1 \rightarrow
b_1 \pi$ is allowed whereas a hybrid $1^{++}$ is forbidden in this channel.

These various signals in the desired channels provide a potentially consistent
picture. The challenge now is to test it. Dedicated high statistics
experiments with the power of modern detection and analysis should re-examine
these channels.
As several of these candidates appear in photoproduction, it may be worthwhile
to consider the possibility that the ($L=0$) + ($L=0$) suppression
(into $\rho \omega$ or $\rho \pi$) may be overruled when the
``$\rho$" is a $\gamma$. With this in mind we
identify from tables 7-9 of ref.\cite{cp94} the potentially significant
photon couplings (where ``photon" is equated with $\rho$ {\it via} vector meson
dominance and the exchanged
particle is $\omega$ or $\pi$ in this analysis).

The unsuppressed widths are for I=1 hybrid states (in keV)

\beqna
2^{-+}  \rightarrow  \hsp \pi^+ \gamma = & \hst \; \: 70;
& \omega \gamma = 250 \\ \nonumber
1^{-+}  \rightarrow  \hsp \pi^+ \gamma = & \hst \; \: 70;
& \omega \gamma = 180 \\ \nonumber
0^{-+}  \rightarrow  \hsp \pi^+ \gamma = & \hst 275; &  \omega \gamma
=   \;\; \; \,  0 \\ \nonumber
1^{++}  \rightarrow  \hsp \pi^+ \gamma = & \hst 145; & \omega \gamma = 455
\eeqna

and for the $I=0$ hybrid sector important widths include

\beqna
2^{-+}  \rightarrow  \hsp \rho \gamma & = & 127 \\ \nonumber
1^{-+}  \rightarrow  \hsp \rho \gamma & = & \; \:  90 \\ \nonumber
1^{++}  \rightarrow  \hsp \rho \gamma & = & 220 \\ \nonumber
1^{--}  \rightarrow  \hsp \pi  \gamma & = & 120 \\ \nonumber
\eeqna

The $1^{++}$ channel has the strongest photon couplings but is predicted
to be broad and hard to isolate (unless its mass is unexpectedly
low, as for the enigmatic $f_1(1430)$ in the isoscalar sector).
 Both $1^{-+}, 2^{-+}$ have
similar photoproduction rates and healthy $b_1 \pi$ branching ratios and hence
are candidates for the signal in ref.\cite{atkinson}. The $0^{-+}$
may be prominent in low energy photoproduction where $\pi$ exchange is
important but its $\omega \gamma$
coupling is suppressed by a selection rule for hybrids which will disfavour
the $0^{-+}$ photoproduction at higher energies.

The above are all upper limits on the electromagnetic couplings and may
be used to estimate upper limits for the photoproduction rates. In the case
of $\pi$ exchange, at least, there is reason to expect that the actual
strengths will be $\geq 20\%$ of these\cite{cp94}.
 This is especially favourable to
{\it low-energy} photoproduction and as such offers a rich opportunity
for the programme at an upgraded CEBAF
 or possibly even at HERA. If the results of ref.\cite{atkinson}
are a guide, then photoproduction may be an important gateway at a range of
energies and the channel $\gamma + N \rightarrow (b_1 \pi) + N$ can
discriminate
hybrid $1^{--}$ and $2^{-+}$ from their conventional counterparts.

We also note that in $\pi^- p \rightarrow (f_1 \pi^-) p$ ref.\cite{lee94}
sees a broad $1^{++}$ signal that is more prominent than their exotic $1^{-+}$
candidate. They interpret this $1^{++}$ as a radial
excitation; however, we note
that if the production mechanism involves $\rho$ exchange, then the hybrid
couplings predicted in ref.\cite{cp94} imply that hybrid $1^{++}$ should be
produced in this same experiment with a greater strength than the $1^{-+}$.
Experimentally it may be possible to distinguish between the radial $^3P_1$
axial meson and the hybrid (spin singlet) configuration by studying
$\pi N$ or $\gamma N \rightarrow (b_1 \pi) N$ : a hybrid $1^{++}$
decouples from this channel.

Thus to summarise, we suggest that data are consistent with the existence of
low lying multiplets of hybrid mesons based on the mass spectroscopic
predictions of ref.\cite{paton85} and the production and decay dynamics of ref.
\cite{cp94}. Specifically the data include

\beqna
 0^{-+} & (1790 \; MeV; \Gamma = 200 \; MeV) & \rightarrow
\hsp \pi f_0 ; K\bar{K}\pi \\
\nonumber
 1^{-+} & (\sim 2 \; GeV; \Gamma \sim 300 \;  MeV)  & \rightarrow
\hsp  \pi f_1 ; \pi b_1 (?) \\
\nonumber
 2^{-+} & (\sim 1.8 \; GeV; \Gamma \sim  200 \; MeV)  & \rightarrow
\hsp  \pi b_1; \pi f_2 \\
\nonumber
 1^{--} & (1460 \; MeV; \Gamma \sim 300 \; MeV)  & \rightarrow
\hsp  \pi a_1
\eeqna

Detailed studies of these and other relevant channels are called for together
with analogous searches for their hybrid charmonium analogues, especially in
photoproduction or $e^+ e^-$ annihilation.

\vskip 0.2in
\noindent {\bf Acknowledgements}

We thank T. Barnes, A. Donnachie and several experimental colleagues
at the Rutherford Appleton Laboratory for helpful discussions.

\end{document}